\documentclass[conference]{IEEEtran}
\IEEEoverridecommandlockouts
\usepackage{cite}
\usepackage{amsmath,amssymb,amsfonts}
\usepackage{algorithmic}
\usepackage{graphicx}
\usepackage{textcomp}
\usepackage{xcolor}
\def\BibTeX{{\rm B\kern-.05em{\sc i\kern-.025em b}\kern-.08em
    T\kern-.1667em\lower.7ex\hbox{E}\kern-.125emX}}

\newcommand{\DeepGCC}{\texttt{DeepGCC}}
\newcommand{\ASLNet}{\texttt{ASLNet}}
\newcommand{\RecuNet}{\texttt{SELnetXYZ}}
\newcommand{\SRP}{\texttt{SRP-PHAT}}

\newcommand{\SELDnet}{\texttt{SELDnet}}

\usepackage{subfigure}
\usepackage{multirow}
\graphicspath{{./}{./figures/}{./images/}}
\begin{document}

\title{Towards Domain Independence in CNN-based Acoustic Localization
  using Deep
  Cross Correlations\\
  \thanks{This work has been partially supported by the Spanish Ministry
    of Economy and Competitiveness under projects HEIMDAL-UAH
    (TIN2016-75982-C2-1-R) and ARTEMISA (TIN2016-80939-R); and by the
    University of Alcal\'a under project ACERCA (CCG2018/EXP-019).}  }

\author{\IEEEauthorblockN{Juan Manuel Vera-Diaz}
  \IEEEauthorblockA{\textit{Department of Electronics} \\
    \textit{University of Alcala.}\\
    Alcala de Henares, Madrid\\
    ORCID 0000-0002-6152-5789}
  \and
  \IEEEauthorblockN{Daniel Pizarro}
  \IEEEauthorblockA{\textit{Department of Electronics} \\
    \textit{University of Alcala.}\\
    Alcala de Henares, Madrid\\
    ORCID 0000-0003-0622-4884}
  \and
  \IEEEauthorblockN{Javier~Macias-Guarasa}
  \IEEEauthorblockA{\textit{Department of Electronics} \\
    \textit{University of Alcala.}\\
    Alcala de Henares, Madrid\\
    ORCID 0000-0002-3303-3963}
}

% \author{\IEEEauthorblockN{Frank~Sanabria-Macias, Marta~Marron-Romera and Javier~Macias-Guarasa}
% \IEEEauthorblockA{\textit{GEINTRA Research Group, Department of Electronics} \\
% \textit{University of Alcala.}\\
% Alcala de Henares, Madrid\\
% frank.sanabria@edu.uah.es, Marta~Marron-Romera, Javier~Macias-Guarasa}
% }

\maketitle

\begin{abstract}
  Time delay estimation is essential in \textit{Acoustic Source
    Localization} (ASL) systems. One of the most used techniques for
  this purpose is the \textit{Generalized Cross Correlation} (GCC)
  between a pair of signals and its use in \textit{Steered Response
    Power} (SRP) techniques, which estimate the acoustic power at a
  specific location. Nowadays, \textit{Deep Learning} strategies may
  outperform these methods. However, they are generally dependent on the
  geometric and sensor configuration conditions that are available
  during the training phases, thus having limited generalization
  capabilities when facing new environments if no re-training nor
  adaptation is applied. In this work, we propose a method based on an
  encoder-decoder \textit{CNN} architecture capable of outperforming the
  well known \textit{SRP-PHAT} algorithm, and also other \textit{Deep
    Learning} strategies when working in mismatched training-testing
  conditions without requiring a model re-training. Our proposal aims to
  estimate a smoothed version of the correlation signals, that is then
  used to generate a refined acoustic power map, which leads to better
  performance on the ASL task. Our experimental evaluation uses three
  publicly available realistic datasets and provides a comparison with
  the \textit{SRP-PHAT} algorithm and other recent proposals based on
  \textit{Deep Learning}.
\end{abstract}


\section{Introduction}
\label{sec:intro}

One of the critical tasks in \textit{Acoustic Source Localization (ASL)}
is the time delay estimation~\cite{Brandstein97practical, huang2008time}
between signals recorded by a pair of acoustic sensors that are
generated by an unknown acoustic source. With at least three of these
pairs, it is possible to estimate the position of the acoustic source by
using hyperbolic trilateration techniques. However, this process is not
reliable in everyday scenarios with signals contaminated with noise and
multipath effects.  Other \textit{ASL} methods are more robust to these
effects such as those based on the \textit{Steered Response Power
  (SRP)}~\cite{dibiase2000high, dibiase2001robust, Dmochowski2007,
  Wan2010ImprovedSRP, Hoang2010} or the \textit{Minimum Variance
  Distortionless Response (MVDR)}~\cite{habets2010MVDR, salvati2016};
all of them based on the Generalized Cross Correlation
(GCC)~\cite{knapp1976GCC}.

% One of the key tasks in \textit{Acoustic Source Localization (ASL)} is
% the time-delay estimation \cite{Brandstein97practical, huang2008time}
% between signals recorded by a pair of acoustic sensors that are
% generated by an unknown acoustic source. Given at least three of these
% pairs, it is possible to estimate the position of the acoustic source by
% using hyperbolic trilateration techniques. However, this process is not
% reliable in common scenarios where the signals are contaminated with
% noise and multipath effects. Other \textit{ASL} methods are more robust
% to these effects such as those based on the \textit{Steered Response
% Power} (SRP)~\cite{dibiase2000high, dibiase2001robust, Dmochowski2007,
% Wan2010ImprovedSRP, Hoang2010} or the Minimum Variance Distortionless
% Response (MVDR)~\cite{habets2010MVDR, salvati2016}, %The Generalized
%                                 % Cross Correlation (GCC)
%                                 % \cite{knapp1976GCC} is a common
%                                 % component.  
% all of them based Generalized Cross Correlation
% (GCC)~\cite{knapp1976GCC}.

In recent years, \textit{ASL} methods based on \textit{Deep Learning} techniques have 
appeared in the literature. In~\cite{Vera2018Towards}, they use raw acoustic signals to 
estimate the source position coordinates directly, and in~\cite{adavanne2018}, they use 
the signal spectra to estimate the \textit{Direction of Arrival (DoA)} of the 
acoustic source. Their results are promising, reporting better accuracy 
than classical methods. However, they have significant limitations: 
\textit{1)} they require a large amount of labeled data for training, whereas there 
is limited availability of large and public datasets for \textit{ASL}, and \textit{2)} 
learning techniques are highly dependent on the room and sensor
geometry, and the training 
conditions. As a consequence, their accuracy dramatically degrades when used 
in other environments or outside the physical area used to generate training examples.

In this paper, we propose a method based on a \textit{Convolutional
  Neural Network (CNN)}. It takes the \textit{GCC} of a pair of signals
in its input and estimates a likelihood function where its maximum
appears at the time-delay between the two signals. We then combine these
likelihood functions in a 3D spatial grid, as proposed in classical
\textit{SRP} techniques. The target is similar to that described
in~\cite{Hougnigan2017Neural}, in which they propose a method to obtain
the time-delay between two signals using the GCC and multilayer
perceptrons with a single hidden layer. However, their system is only
tested using artificial signals (chirps), so that it is not possible to
assess its applicability in realistic scenarios.

Our contributions are the following: \textit{1)} our method is largely
independent of the room and sensor geometry, \textit{2)} we can use small size datasets to train the neural network and \textit{3)} it is consistently
more accurate than classical methods such as \textit{SRP-PHAT} and also
better than other \textit{Deep Learning} methods in general conditions, where the 
testing room is physically different, or the source position
significantly differs from those available in the training data.

\section{Problem Statement}
\label{sec:ProblemStatement}

Let us consider an environment where we place $M$ microphones at known
positions $\vec{m}_k = (m_{x_k},m_{y_k},m_{z_k})^\top$ with $k = 0,
\dots, M-1$. An acoustic source emits a signal $s(t)$ which is received
and sampled by each microphone, obtaining a discrete-time signal
$x_k[n]$ at the $k^{th}$ microphone. % and where $f_s$ is the sampling
% rate.   
% $x_k[n] = h_k(\frac{n}{f_s}) * s(\frac{n}{f_s}) + v[n]$ where $f_s$ is the sampling rate, $*$ is the convolution operator, $h_k$ is the \textit{Room Impulse Response (RIR)} between the source and the microphone and $v$ is the additive noise. From these $M$ received signals, the \textit{Acoustic Power (AP)} at an specific location denoted by $\vec{q} = (q_x, q_y, q_z)^\top$  can be calculated by the \textit{SRP-PHAT} beamformer as is shown in equation \ref{eq:eq1}.

% \begin{equation}
%   AP(\vec{q}) = \sum_{k=0}^{M-1} \sum_{l=k+1}^{M-1} \mathcal{F}^{-1}\left( \dfrac{X_k \cdot X_l^*}{\mid X_k \mid \cdot \mid X_l \mid} \right)[\Delta \tau_{(\vec{q}, \vec{m}_k, \vec{m}_l)} \, f_s]
%   \label{eq:eq1}
% \end{equation} 

% With $X_k$ as the \textit{DFT} of signal $x_k[n]$, $\cdot$ as the
% element-wise product operator, $(\;)^*$ as the conjugate operator,
% $\mid (\;) \mid$ as the magnitude operator, $\mathcal{F}^{-1}$ as the
% inverse \textit{DFT} and $\Delta \tau_{(\vec{q}, \vec{m}_k,
% \vec{m}_l)}$ as the time-delay between the signals received by the
% microphone pair $k$-$l$ according to the position $\vec{q}$ which is
% defined by the equation \ref{eq:eq2}.

In fully ideal conditions the signals received at microphones $k$ and
$l$ from a source at position $\vec{q} = (q_x, q_y, q_z)^\top$ only
differ in a time delay $\Delta \tau_{(\vec{q}, \vec{m}_k, \vec{m}_l)}$:
\begin{equation}
  \Delta \tau_{(\vec{q}, \vec{m}_k, \vec{m}_l)} = \dfrac{\parallel \vec{q} - \vec{m}_k \parallel - \parallel \vec{q} - \vec{m}_l \parallel}{c},
  \label{eq:eq2} 
\end{equation}
where $\parallel . \parallel$ represents the Euclidean norm and $c$ is
the sound propagation velocity ($340 \; m/s$ at $20 \; ^{\circ}C$). 
The \textit{GCC-PHAT} between the two signals is defined as:
\begin{equation}
  gcc(x_k[n],x_l[n]) = \mathcal{F}^{-1}\left( \dfrac{X_k \cdot X_l^*}{\mid X_k \mid \cdot \mid X_l \mid} \right),
  \label{eq:corr} 
\end{equation}
where $X_k$ is the \textit{DFT} of signal $x_k[n]$, $\cdot$ is the
element-wise product operator, $(\;)^*$ is the conjugate operator, $\mid
. \mid$ is the magnitude and $\mathcal{F}^{-1}$ denotes the inverse
\textit{DFT}.  
% The \textit{Acoustic Power (AP)} at an specific location denoted by $\vec{q} = (q_x, q_y, q_z)^\top$  can be calculated by the \textit{SRP-PHAT} beamformer as is shown in equation \ref{eq:eq1}.
% According to equation \ref{eq:eq1}, the main term on it is the one inside the inverse \textit{DFT} operator, which is the \textit{GCC} of a pair of signals in the frequency domain. 
% It can be shown that if one of the signals is a shifted version of the other, and there is no presence of any other type of distortion on them, the signals
Under ideal propagation conditions, the \textit{GCC-PHAT} is a Dirac's
delta shifted according to the time-delay between the signals. In a real
scenario, the microphone signals are affected by the distortion
and reverberation introduced by the environment, and the
GCC-PHAT signals do not easily allow for the recovery of accurate time delays.
A common approach to overcome this issue is to compute
the so-called \textit{Acoustic Power Map} (APM), evaluated on a grid of
possible source positions $\vec{q}_0,\dots,\vec{q}_K$, by using the
\textit{SRP-PHAT} beamformer:
\begin{equation}
  APM(\vec{q}) = \sum_{k=0}^{M-1} \sum_{l=k+1}^{M-1} gcc\left(x_k[n],x_l[n]\right)% [\Delta \tau_{(\vec{q}, \vec{m}_k, \vec{m}_l)} \, f_s]
  \label{eq:eq1}
\end{equation} 
The source position that maximizes $APM$ is an estimate of the true source position. 
% The position where the maximum of the $APM$ map is obtained is used as a
% estimation of the true source position.  

Our proposal is a encoder-decoder \textit{CNN}, represented by the
mathematical function $\mathit{f}_{net}$, that takes as input the
\textit{GCC-PHAT} between two signals and produce a
% That defines our goal, to find a function capable of giving a simplified and smoothed version of the \textit{GCC} output. Specifically, this function will give a 
Gaussian-like signal with variance $\sigma^2$ and mean equal to the
time-delay shift (see Figure~\ref{fig:figure1}):% This function is shown
% in equation \ref{eq:eq3} and will be
% modeled using a \textit{CNN}. 
\begin{eqnarray}
  \mathit{f}_{net}(gcc(x_k[n],x_l[n])) = e^{\dfrac{-(D - \Delta \tau_{(\vec{q}, \vec{m}_k, \vec{m}_l)} \, f_s)^2}{2*\sigma^2}} \nonumber\\
  \label{eq:eq3} 
\end{eqnarray}
with $D=-L/2,\dots,L/2$, and $L$ being the maximum possible sample delay
according to the microphone topology. We can define the \textit{APM}
based on our method by using equation~\eqref{eq:eq3} as an estimation of
the $gcc$ function in equation~\eqref{eq:eq1}. Our
method produces a smoother $APM$ (see Figure~\ref{fig:figure3}), and yields
better \textit{ASL} performance than \textit{SRP-PHAT}, as described in
section~\ref{sec:Results}.

\section{Proposed CNN Architecture}
\label{sec:ProposedModel}

\begin{figure}[htb]
	\centering
	\includegraphics[width=\columnwidth]{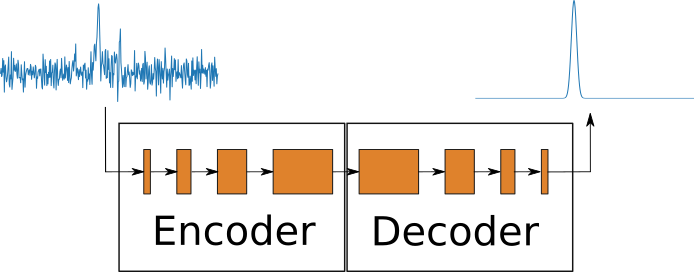}
	\caption{\DeepGCC{} layers scheme.}
	\label{fig:figure1}
\end{figure}

The proposed CNN model that implements equation~\eqref{eq:eq3}, from now on \DeepGCC{}, 
is shown in Figure~\ref{fig:figure1}. 
% The proposed \textit{CNN} model, hereinafter \DeepGCC{}, that implements
% equation \ref{eq:eq3} is shown in Figure~\ref{fig:figure1}.% This architecture takes as input the
% \textit{GCC} of a pair of signals and
% outputs . 
\DeepGCC{} uses an encoder-decoder architecture with $36025$ parameters. 
Both the encoder and the decoder use four blocks composed of a 1D convolutional 
layer with kernels of size $4$, Max-Pooling (encoder) or upsampling (decoder) layers with size of two samples,
batch-normalization, and ReLU activations. Table~\ref{tab:table1}  summarizes the input and output sizes of every block.

% The depth of the convolutional layer of
% each block is 2, 8, 32 and 128 respectively.  The decoder is formed by
% four blocks, composed each by an upsampling layer of size two, a 1D
% convolutional layer with $4$ coefficient kernel, a batch-normalization
% layer and a ReLU activation in all layers except the last one, in which
% a sigmoid activation is used. The depth of the convolutional layer of
% each block is 32, 8 and 2, 1 respectively. 
\begin{table}[htb]
	\centering
  \small
	\begin{tabular}{| c | c | c |}

		\hline
		\textbf{Block} & \textbf{Input size} & \textbf{Output size}\\

		\hline \hline
		Encoder: Block 1 & $L \times 1$ & $L/2 \times 2$ \\
		\hline
		Encoder: Block 2 & $L/2 \times 2$ & $L/4 \times 8$ \\
		\hline
		Encoder: Block 3 & $L/4 \times 8$ & $L/8 \times 32$ \\
		\hline
		Encoder: Block 4 & $L/8 \times 32$ & $L/16 \times 128$ \\

		\hline \hline
		Decoder: Block 1 & $L/16 \times 128$ & $L/8 \times 32$ \\
		\hline
		Decoder: Block 2 & $L/8 \times 32$ & $L/4 \times 8$ \\
		\hline
		Decoder: Block 3 & $L/4 \times 8$ & $L/2 \times 2$ \\
		\hline
		Decoder: Block 4 & $L/2 \times 2$ & $L \times 1$ \\
		\hline
	\end{tabular}
	\caption{Summary of the input and output sizes of each network block.}
	\label{tab:table1}
\end{table}

% \TODO{Check esto, hay que justificar con detalles y ponerlo en parte experimental}
% In our experiments we use $L=400$ as it is a suitable value given the
% sampling rate and the maximum time-delay. This is later detailed in
% section \ref{sec:DG}.

\begin{figure*}[htb]
	\centering
	\includegraphics[width=0.8\textwidth]{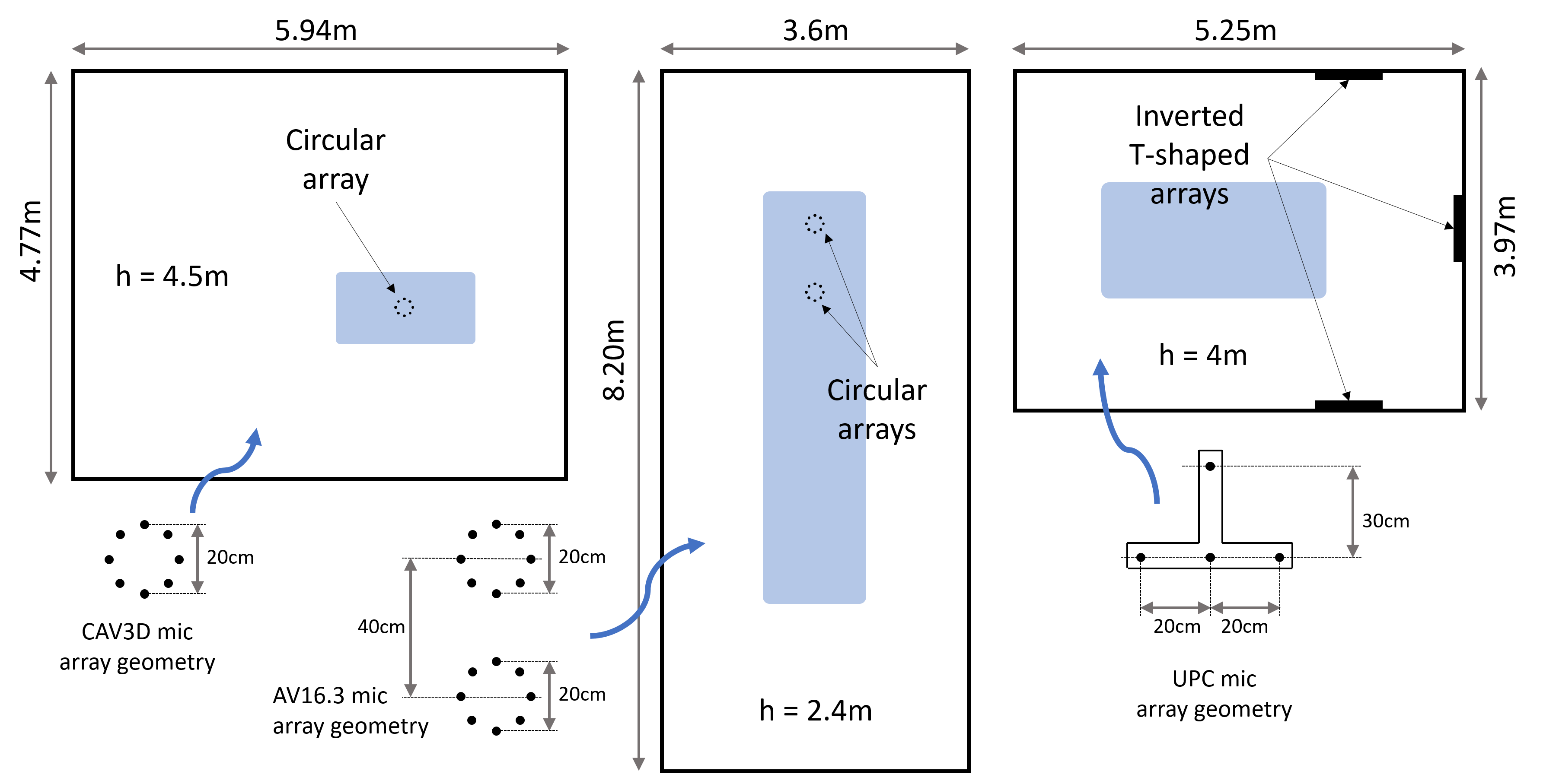}
	\caption{Top view of the rooms used: CAV3D room (left), AV16.3 (center) and UPC (right).}
	\label{fig:room-geometries}
\end{figure*}

\section{Experimental Setup}
\label{sec:experimental-setup}

\subsection{Datasets}
\label{UD}

In order to evaluate till what extent the proposed method works properly
under different acoustic and geometric conditions, we used three
different datasets for the training and testing phases (refer to
Figure~\ref{fig:room-geometries} for graphical details):

% as our main objective of this approach is
% to develop a \textit{deep learning} method capable of properly working
% in different scenarios without retraining. 
\begin{itemize}
\item The \textit{CAV3D} dataset~\cite{Qian2019Multi} was recorded at
  $96kHz$ in a rectangular room of $4.77 m \times 5.94 m \times 4.50m$,
  with a circular array of $8$ microphones and $10 cm$ radius, placed on
  top of a $73 cm$ height table. All acoustic frames include the
  speaker's 3D mouth position coordinates. This dataset is composed of
  $10$ single and moving speaker sequences, with varying user
  characteristics and moving patterns.

\item The \textit{AV16.3} dataset~\cite{lathoud2005av16} was recorded in
  a $3.6m \times 8.2m \times 2.4m$ rectangular room that includes two
  circular arrays with the same geometry as the \textit{CAV3D} array,
  also placed on top a $73cm$ height table. This paper focuses on a
  subset of the dataset composed of $5$ single speaker sequences
  recorded at $16kHz$, comprising three static user and two moving user
  sequences. The recordings also include varying user characteristics
  and moving patterns. In our experimental work, we only use one
  circular array and upsample the signal to $96Khz$, to provide the same
  array configuration and sampling rate as those of the \textit{CAV3D}
  dataset.

\item The \textit{CHIL-CLEAR} dataset~\cite{stiefelhagen2007clear} contains sequences of ``lecture seminars'' with $3$ to $8$ participants
  (with no overlapped speech) recorded at five different rooms
  (\textit{AIT}, \textit{IBM}, \textit{ITC}, \textit{UKA}, and
  \textit{UPC}). In each room, eight sequences of $5$ minutes were
  recorded. In this work, we focus only on the $8$ \textit{UPC} sequences,
  which were recorded in a $3.97m \times 5.25m \times 4.00 m$ room,
  using three inverted T shaped 4-microphone arrays located on the walls
  at the height of $2.38m$. Again, we upsampled the microphone signals up
  to $96kHz$.

\end{itemize}

Figure~\ref{fig:room-geometries} shows the top view of the three rooms
corresponding to the CAV3D (left), AV16.3 (center), and UPC (right),
adequately scaled to provide the reader with a visual clue on the
different geometrical and acoustical conditions for each database. It also shows the
specific geometry of the microphone array configurations for each room.

\begin{table}[htb]
	\centering
	  \resizebox{\columnwidth}{!}{
%\footnotesize
	\begin{tabular}{| c | c | c || c | c | c || c | c | c |}
	\hline
	\textbf{\textit{CAV3D}} & \textbf{\#} & \textbf{Charact.} & \textbf{\textit{AV16.3}} & \textbf{\#}  & \textbf{Charact.} & \textbf{\textit{UPC}} & \textbf{\#}  & \textbf{Charact.}\\
	\hline \hline
	\texttt{C06} & 393 & $M$     & \texttt{A01}  & 1328 & $S+3$ & \texttt{U01} & 3869 & $S+3$\\
	\hline
	\texttt{C07} & 374 & $M$     & \texttt{A02}  & 1441 & $S+3$ & \texttt{U02} & 4455 & $S+3$\\
	\hline
	\texttt{C08} & 412 & $M$     & \texttt{A03}  & 1552 & $S+3$ & \texttt{U03} & 3794 & $S+3$\\
	\hline
	\texttt{C09} & 367 & $M+1$   & \texttt{A11}  &  297 & $M+3$ & \texttt{U04} & 3780 & $S+3$\\
	\hline
	\texttt{C10} & 245 & $M+1$   & \texttt{A15}  &  284 & $M+3$ & \texttt{U05} & 4348 & $S+3$\\
	\hline
	\texttt{C11} & 709 & $M+1+2$ & ---              &  --- &   --- & \texttt{U06} & 3915 & $S+3$\\
	\hline
	\texttt{C12} & 725 & $M+1+2$ &             ---  &  --- &   --- & \texttt{U07} & 3135 & $S+3$\\
	\hline
	\texttt{C13} & 684 & $M+3$   &             ---  &  --- &   --- & \texttt{U08} & 3734 & $S+3$\\
	\hline
	\texttt{C20} & 422 & $M$     &             ---  &  --- &   --- & ---             &  --- &   ---\\
	\hline
	\texttt{C21} & 476 & $M+2$   &             ---  &  --- &   --- & ---             &  --- &   ---\\
	\hline
	\end{tabular}
}
\caption{\textit{CAV3D}, \textit{AV16.3}, \textit{UPC} sequences used
  to train and test our method. The second, fifth and eighth columns include
  the number of available acoustic frames in the sequence, and the
  third, sixth and last columns represent the characteristics of each sequence.}
	\label{tab:table2}
\end{table}

Table~\ref{tab:table2} summarizes the characteristics of the used
sequences.  We selected the sequences from \textit{CAV3D} for training,
validation, and testing, and the sequences from \textit{AV16.3} and
\textit{UPC} for testing (see section~\ref{sec:experimental-design} for
details). We code the characteristics of each sequence as follows:
$S$ refers to static speakers, $M$ refers to moving speakers, $1$ denotes
noise present in the sequence, $2$ denotes a speaker at two different
heights, and $3$ refers to the case of target positions not present in
the rest of the sequences (thus not available in the training stages).

\subsection{Dataset Processing}
\label{sec:DG}

The process of generating the training, validation, and testing subsets is
identical in all cases and consists of computing the \textit{GCC-PHAT}
for all the possible signals pairs. We extract each acoustic frame
from the whole sequence using a $166 ms$ signal frame with $50\%$
overlapping and Blackman windowing. Assuming a $96kHz$ sampling rate, we
use windows with $400$ samples, which implies that the network can
process time-delays up to approximately $\pm 2 ms$. This delay is more
than enough, given the maximum separation between microphones in the
training data ($20 cm$, as we only use the \textit{CAV3D} dataset for training).

For each \textit{GCC-PHAT} signal, we generate the supervised network output according to equation~\eqref{eq:eq3}, computing the time-delay between the signals received by the two microphones and the labeled source position. This signal is also generated to be $400$
samples in length and has a standard deviation of $\sigma = 5$ samples
that we empirically selected in preliminary experiments.

\subsection{Training Procedure}
\label{sec:TP}

For the training phase, we followed the same procedure to evaluate all
the methods. The loss function consists of the \textit{Mean Squared
  Error (MSE)} between the network output and the target signal, generated with
equation~\eqref{eq:eq3} using the \textit{ground truth} source
position. To minimize the loss, we used the \textit{Adam} optimizer
\cite{adam2014} with a \textit{learning rate} of $10^{-4}$ and a
\textit{decay} of $10^{-8}$, leaving the rest of the parameters at their
default values.  Batch size is equal to $100$ samples, and we used
validation data to stop training if the loss does not improve during
$50$ consecutive epochs.
      %       Each epoch data is input to the network in batches of 100 samples and the stop criteria used to finish the training is to not getting a better validation data loss value in .
The \textit{CAV3D} dataset was the only used for training, and due to the
different features of each sequence, we run three different partitions,
as shown in Table~\ref{tab:table3}, leaving the hardest sequences
(\texttt{C09}, \texttt{C10}, \texttt{C11}, \texttt{C12}, and especially
\texttt{C13} (that contains speaker positions not available in the
training subset)) for validation and testing, and using the other simpler
ones for training.

\begin{table}[htb]
	\centering
  \footnotesize
	\begin{tabular}{| c | c | c | c |}
    \hline
    \textbf{Partition \#} & \textbf{Test} & \textbf{Val} & \textbf{Train}\\
    \hline \hline
    P1 & \texttt{C10}, \texttt{C12} & \texttt{C13} & Rest of \textit{CAV3D} sequences \\
    \hline
    P2 & \texttt{C09}, \texttt{C11} & \texttt{C13} & Rest of \textit{CAV3D} sequences  \\
    \hline
    P3 & \texttt{C13} & \texttt{C11} & Rest of \textit{CAV3D} sequences  \\
    \hline
	\end{tabular}
	\caption{Evaluated training/validation/testing partitions.}
	\label{tab:table3}
\end{table}

\subsection{Algorithms Comparison}
\label{sec:ED}

      %       \subsection{Other methods}

To compare the accuracy of our proposal, we evaluate it against three
alternatives. The first one is the well known \SRP{} algorithm,
considered as the baseline system. The second method is the \textit{Deep
  Learning} approach \ASLNet{}~\cite{Vera2018Towards}, which estimates
the cartesian coordinates of the acoustic source from the raw audio
signals of a set of microphones. The last evaluated method is based on
\SELDnet{}~\cite{adavanne2018}, a recurrent neural network to estimate
the azimuth and elevation of an acoustic source from the spectrogram of
the audio signal of a microphone array. In our work, we have modified
this architecture to estimate the source 3D cartesian coordinates
directly, and we will refer to it as \RecuNet{}.

\subsection{Experimental design}
\label{sec:experimental-design}

We carried out two different experiments to focus on two performance
indicators: ASL precision performance, and environmental robustness:

\begin{itemize}
\item For testing the ASL precision performance, we used the
  \texttt{C09}, \texttt{C10}, \texttt{C11}, and \texttt{C12} sequences
  because of the similarity of its labeled positions with the
  training ones.  In this experiment, we expect better performance for
  the standard \textit{Deep Learning} approaches \ASLNet{} and
  \RecuNet{}.  We also expect that our proposal will achieve better
  performance than the \SRP{} beamformer.  We evaluated the ASL
  performance as the mean Euclidean distance between the labeled
  \textit{ground truth} and the estimated position, provided the
  training and testing positions are similar.

\item For testing the environmental robustness of the different
  proposals, we measured the localization accuracy when the room
  geometry, microphone array geometry, and the evaluated positions in the
  test data significantly differ from those in the training data.
  \textit{IDIAP} and \textit{UPC} sequences were used for this purpose,
      %       Sequences \texttt{A01}, \texttt{A02}, \texttt{A03}, \texttt{A11} and \texttt{A15} are used for this purpose, 
  along with \texttt{C13}, which belong to the \textit{CAV3D} dataset,
  but includes a full range of positions not found in the training
  subsets. In this case, we expect our \DeepGCC{} proposal to roughly
  keep the same performance as in the first experiment, while the other
  \textit{Deep Learning} methods exhibit a performance
  decrease due to mismatched evaluation conditions. Note that in the
  \textit{UPC} sequences, only the \SRP{} and \DeepGCC{} algorithms can
  be evaluated since the microphone array geometry has changed, and we
  will not retrain \ASLNet{} or \RecuNet{}.

%. The other \textit{Deep Learning} proposals
%  are not available to be evaluated on \textit{UPC} as their trained
%  models are not consistent with the array geometry.

\end{itemize}
      %       Due to the short ammount of data of the \textit{CAV3D} dataset (table
      %       \ref{tab:table2}) we have decied to train three times with differents
      %       sequences in first experiment in order to have more testing
      %       sequences. This split is shown in table \ref{tab:table3}. In the other
      %       hand, we heve use all the single speaker sequences of the
      %       \textit{AV16.3} database in experiment two, where niether a re-training
      %       or a fine-tuning process has been carry out for any model tested.

      %       \begin{table}[htb]
      %       \centering
      %       \begin{tabular}{| c | c | c | c |}
      %       \hline
      %       \textbf{Training \#} & \textbf{Test} & \textbf{Val} & \textbf{Train}\\
      %       \hline \hline
      %       Training 1 & $10$, $12$ & $13$ & Rest of them \\
      %       \hline
      %       Training 2 & $seq09$, $seq11$ & $seq13$ & Rest of them\\
      %       \hline
      %       Training 3 & $seq13$ & $seq11$ & Rest of them\\
      %       \hline
      %     \end{tabular}
      %       \caption{Sequences separation for the first experiment proposed.}
      %       \label{tab:table3}
      %       \end{table}

In all the experiments, we have used all the possible microphone pair
combinations to build a volumetric acoustic power map with a
$10cm \times 10cm \times 10cm$ grid resolution, generated from the \DeepGCC{} model output. 
We then extract the acoustic source cartesian coordinates from the location
of the maximum acoustic power. 

\section{Results}
\label{sec:Results}

\begin{figure}[htb]
	\centering
	% \begin{tabular}{c c}
      %      \includegraphics[width=0.2\textwidth]{InputOutputExample} &
                                                                         %                                                                         \includegraphics[width=0.2\textwidth]{Occupation2}\\
      %    \end{tabular}
      %       \includegraphics[height=2.5cm]{All2gether}
	\includegraphics[width=\columnwidth]{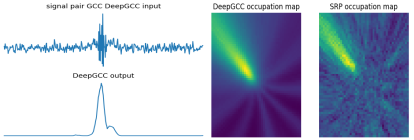}
	\caption{Example of a \textit{GCC-PHAT} input and the \DeepGCC{}
    result and their respective Acoustic Power Maps.}
	\label{fig:figure3}
\end{figure}

Figure~\ref{fig:figure3} (left) shows a particular example of the output
obtained with \DeepGCC{} and \textit{GCC-PHAT} (\SRP{}). Figure~\ref{fig:figure3}
(right) shows the APMs using equation~\eqref{eq:eq1}, which involves all
microphone pairs. The map built with \DeepGCC{} is considerably
smoother than that built with \SRP{}.
      %       After training the \DeepGCC{} model, a sample result is shown in
      %       Figure~\ref{fig:figure3} where the \textit{GCC-PHAT} input and the
      %       network estimation can be compared. If this process is repeated for all
      %       the microphone pairs using the output of the \DeepGCC{} model, the APM
      %       built using equation~(\ref{eq:eq1}) is shown to be smoother than the map
      %       built with the \textit{SRP-PHAT} technique (see also
      %       Figure~\ref{fig:figure3}).
      %       \begin{figure}[htb]
      %       \centering
      %       \includegraphics[height=3cm]{Occupation2}
      %       \caption{Example of a slice of the volumetric occupational map using \textit{GCC-PHAT} as input to the %\textit{SRP} beamformer and using \DeepGCC{} output as \textit{SRP} beamformer input.}
      %       \label{fig:figure4}
      %       \end{figure}

      %       Once we can build up a volumetric accupational map, we can estimate the
      %       location of the acoustic source by searching for the maximum value in
      %       the volumen. The error between this estimated position and its labeled
      %       \textit{ground truth} is measured in terms of \textit{MSE} in $cm$. Also
      %       this estiamtion error is obtained for the \textit{SRP-PHAT} beamformer
      %       and the \textit{deep learning} models \ASLNet{}
      %       \cite{Vera2018Towards} and \RecuNet{} which is a recurrent
      %       architecture network based on the approach developed in
      %       \cite{adavanne2018} but modified to work in this specific task.

Table~\ref{tab:table5} shows the results of the first experiment
(focused on ASL performance). The average MSE is shown in cm, and
we also include the relative improvement achieved by each method as
compared with the \SRP{} algorithm
($\Delta_r^{MSE}=\frac{MSE_{\SRP}-MSE_{algorithm}}{MSE_{\SRP}}$).
\RecuNet{} obtains, as expected, the best results because testing and
training positions are similar.
      %       Sequence \texttt{C13} is the
      %       only one where appears positions outside the training area, and is in
      %       this sequence where our proposal gets the best scores. 
Our \DeepGCC{} proposal is the second-best strategy, getting better
results than \SRP{} in all cases, and better than \ASLNet{} in 3 out of 4
sequences.
\begin{table}[htb]
	\centering
  \resizebox{\columnwidth}{!}{%
      %       \footnotesize
    \begin{tabular}{| c | c | c | c | c | c |}
      \hline
      \textbf{Sequence} & \textbf{\SRP{}} & \textbf{\DeepGCC{}} & \textbf{\ASLNet{}} & \textbf{\RecuNet{}}\\
      \hline \hline
      \begin{tabular}{@{}c@{}}\texttt{C10}\\\scriptsize$\Delta_r^{MSE}$\end{tabular} & \begin{tabular}{@{}c@{}}$101.3$\\\scriptsize$$\end{tabular} & \begin{tabular}{@{}c@{}}$ 94.5$\\\scriptsize$ 6.71\%$\end{tabular} & \begin{tabular}{@{}c@{}}$ 92.5$\\\scriptsize$8.69\%$\end{tabular} & \begin{tabular}{@{}c@{}}$\textbf{55.2}$\\\scriptsize$45.51\%$\end{tabular} \\
      \hline
      \begin{tabular}{@{}c@{}}\texttt{C12}\\\scriptsize$\Delta_r^{MSE}$\end{tabular} & \begin{tabular}{@{}c@{}}$114.5$\\\scriptsize$$\end{tabular} & \begin{tabular}{@{}c@{}}$101.5$\\\scriptsize$11.35\%$\end{tabular} & \begin{tabular}{@{}c@{}}$103.5$\\\scriptsize$9.61\%$\end{tabular} & \begin{tabular}{@{}c@{}}\textbf{73.4}\\\scriptsize$35.89\%$\end{tabular}\\
      \hline
      \begin{tabular}{@{}c@{}}\texttt{C09}\\\scriptsize$\Delta_r^{MSE}$\end{tabular} & \begin{tabular}{@{}c@{}}$94.9$\\\scriptsize$$\end{tabular}  & \begin{tabular}{@{}c@{}}$ 82.1$\\\scriptsize$13.49\%$\end{tabular} & \begin{tabular}{@{}c@{}}$ 87.6$\\\scriptsize$7.69\%$\end{tabular} & \begin{tabular}{@{}c@{}}\textbf{62.0}\\\scriptsize$34.67\%$\end{tabular}\\
      \hline
      \begin{tabular}{@{}c@{}}\texttt{C11}\\\scriptsize$\Delta_r^{MSE}$\end{tabular} & \begin{tabular}{@{}c@{}}$117.5$\\\scriptsize$$\end{tabular} & \begin{tabular}{@{}c@{}}$106.4$\\\scriptsize$ 9.45\%$\end{tabular} & \begin{tabular}{@{}c@{}}$133.3$\\\scriptsize$3.57\%$\end{tabular} & \begin{tabular}{@{}c@{}}\textbf{94.6}\\\scriptsize$19.49\%$\end{tabular}\\
      \hline
      
    \end{tabular}
  }
  \caption{Results for the ASL precision experiments (error in cm, $\Delta_r^{MSE}$ is relative MSE improvement over the \SRP{} algorithm).}  
	\label{tab:table5}
\end{table}  

Table~\ref{tab:table6} shows the results of the second experiment
(focused on environmental robustness), in which our \DeepGCC{} proposal
clearly outperforms all the other methods. The fact that we aim to
estimate the GCC function (that mainly depends on the relative time
delay) makes it more robust to changes in the room and array geometry or
the source positions with respect to the microphone arrays. The other
\textit{deep learning} methods are not able to properly face the
mismatched conditions, performing far worse than the standard \SRP{}. We remark the fact that the \DeepGCC{} proposal has a
$13.82\%$ relative improvement over the \SRP{} algorithm even though the
microphone geometry (\textit{UPC} sequences) changes. This is a difficult
task for a \textit{Deep Learning} method since it has to deal with unseen conditions in the
\textit{GCC-PHAT} signal.

\begin{table}[htb]
	\centering
  \resizebox{\columnwidth}{!}{%
      %       \footnotesize
    \begin{tabular}{| c | c | c | c | c |}
      \hline
      \textbf{Sequence} & \textbf{\SRP{}} & \textbf{\DeepGCC{}} & \textbf{\ASLNet{}} & \textbf{\RecuNet{}} \\
      \hline \hline
      \begin{tabular}{@{}c@{}}\texttt{C13}\\\scriptsize$\Delta_r^{MSE}$\end{tabular} & \begin{tabular}{@{}c@{}}$86.9$\\\scriptsize$$\end{tabular}    & \begin{tabular}{@{}c@{}}\textbf{79.5}\\\scriptsize$8.51\%$\end{tabular} & \begin{tabular}{@{}c@{}}136.7\\\scriptsize-57.31\%\end{tabular} & \begin{tabular}{@{}c@{}}135.5\\\scriptsize-55.53\%\end{tabular}\\
      \hline      \hline
      \begin{tabular}{@{}c@{}}\texttt {A01}\\\scriptsize$\Delta_r^{MSE}$\end{tabular} & \begin{tabular}{@{}c@{}}$103.04$\\\scriptsize$$\end{tabular} & \begin{tabular}{@{}c@{}}\textbf{ 84.4}\\\scriptsize$ 18.09\%$\end{tabular} & \begin{tabular}{@{}c@{}}$348.3$\\\scriptsize$-238.02\%$\end{tabular} & \begin{tabular}{@{}c@{}}$347.8$\\\scriptsize$-237.54\%$\end{tabular}\\
      \hline
      \begin{tabular}{@{}c@{}}\texttt{A02}\\\scriptsize$\Delta_r^{MSE}$\end{tabular} & \begin{tabular}{@{}c@{}}$68.8$\\\scriptsize$$\end{tabular}    & \begin{tabular}{@{}c@{}}\textbf{ 64.6}\\\scriptsize$ 6.10\%$\end{tabular} & \begin{tabular}{@{}c@{}}$350.4$\\\scriptsize$-409.30\%$\end{tabular} & \begin{tabular}{@{}c@{}}$363.2$\\\scriptsize$-427.91\%$\end{tabular}\\
      \hline
      \begin{tabular}{@{}c@{}}\texttt{A03}\\\scriptsize$\Delta_r^{MSE}$\end{tabular} & \begin{tabular}{@{}c@{}}$73.7$\\\scriptsize$$\end{tabular}    & \begin{tabular}{@{}c@{}}\textbf{ 57.7}\\\scriptsize$21.71\%$\end{tabular} & \begin{tabular}{@{}c@{}}$352.3$\\\scriptsize$-378.02\%$\end{tabular} & \begin{tabular}{@{}c@{}}$360.4$\\\scriptsize$ 389.01\%$\end{tabular}\\
      \hline
      \begin{tabular}{@{}c@{}}\texttt{A11}\\\scriptsize$\Delta_r^{MSE}$\end{tabular} & \begin{tabular}{@{}c@{}}$84.1$\\\scriptsize$$\end{tabular}    & \begin{tabular}{@{}c@{}}\textbf{ 69.1}\\\scriptsize$17.84\%$\end{tabular} & \begin{tabular}{@{}c@{}}$274.3$\\\scriptsize$-226.16\%$\end{tabular} & \begin{tabular}{@{}c@{}}$284.1$\\\scriptsize$-237.81\%$\end{tabular}\\
      \hline
      \begin{tabular}{@{}c@{}}\texttt{A15}\\\scriptsize$\Delta_r^{MSE}$\end{tabular} & \begin{tabular}{@{}c@{}}$140.3$\\\scriptsize$$\end{tabular}   & \begin{tabular}{@{}c@{}}\textbf{110.3}\\\scriptsize$21.38\%$\end{tabular} & \begin{tabular}{@{}c@{}}$387.1$\\\scriptsize$-175.91\%$\end{tabular} & \begin{tabular}{@{}c@{}}$386.4$\\\scriptsize$-175.41\%$\end{tabular}\\
      \hline
      \hline
      \begin{tabular}{@{}c@{}}\texttt{U01}\\\scriptsize$\Delta_r^{MSE}$\end{tabular} & \begin{tabular}{@{}c@{}}$96.8$\\\scriptsize$$\end{tabular}    & \begin{tabular}{@{}c@{}}\textbf{ 77.8}\\\scriptsize$19.63\%$\end{tabular} & --- & --- \\
      \hline
      \begin{tabular}{@{}c@{}}\texttt{U02}\\\scriptsize$\Delta_r^{MSE}$\end{tabular} & \begin{tabular}{@{}c@{}}$111.6$\\\scriptsize$$\end{tabular}   & \begin{tabular}{@{}c@{}}\textbf{ 91.9}\\\scriptsize$17.65\%$\end{tabular} & --- & --- \\
      \hline
      \begin{tabular}{@{}c@{}}\texttt{U03}\\\scriptsize$\Delta_r^{MSE}$\end{tabular} & \begin{tabular}{@{}c@{}}$128.8$\\\scriptsize$$\end{tabular}   & \begin{tabular}{@{}c@{}}\textbf{107.5}\\\scriptsize$16.54\%$\end{tabular} & --- & --- \\
      \hline
      \begin{tabular}{@{}c@{}}\texttt{U04}\\\scriptsize$\Delta_r^{MSE}$\end{tabular} & \begin{tabular}{@{}c@{}}$136.2$\\\scriptsize$$\end{tabular}   & \begin{tabular}{@{}c@{}}\textbf{115.3}\\\scriptsize$15.35\%$\end{tabular} & --- & --- \\
      \hline
      \begin{tabular}{@{}c@{}}\texttt{U05}\\\scriptsize$\Delta_r^{MSE}$\end{tabular} & \begin{tabular}{@{}c@{}}$130.0$\\\scriptsize$$\end{tabular}   & \begin{tabular}{@{}c@{}}\textbf{122.1}\\\scriptsize$ 6.08\%$\end{tabular} & --- & --- \\
      \hline
      \begin{tabular}{@{}c@{}}\texttt{U06}\\\scriptsize$\Delta_r^{MSE}$\end{tabular} & \begin{tabular}{@{}c@{}}$130.2$\\\scriptsize$$\end{tabular}   & \begin{tabular}{@{}c@{}}\textbf{113.6}\\\scriptsize$12.75\%$\end{tabular} & --- & --- \\
      \hline
      \begin{tabular}{@{}c@{}}\texttt{U07}\\\scriptsize$\Delta_r^{MSE}$\end{tabular} & \begin{tabular}{@{}c@{}}$103.2$\\\scriptsize$$\end{tabular}   & \begin{tabular}{@{}c@{}}\textbf{ 83.8}\\\scriptsize$18.80\%$\end{tabular} & --- & --- \\
      \hline
      \begin{tabular}{@{}c@{}}\texttt{U08}\\\scriptsize$\Delta_r^{MSE}$\end{tabular} & \begin{tabular}{@{}c@{}}$133.6$\\\scriptsize$$\end{tabular}   & \begin{tabular}{@{}c@{}}\textbf{123.1}\\\scriptsize$ 7.86\%$\end{tabular} & --- & --- \\
      \hline
    \end{tabular}
  }
  \caption{Results for the environmental robustness experiments (error in cm, $\Delta_r^{MSE}$ is relative MSE improvement over the \SRP{} algorithm).} 
	\label{tab:table6}
\end{table}

      %       It is important to note that even all these sequences try to prove the
      %       environmental robustness of the network, \texttt{C13} proves it
      %       through the accuracy obtained in different positions from the training
      %       but in the same room while all of the other sequences changes the whole
      %       scenario and the positions. This is the reason classic \textit{deep
      %       learning} approach gets lower errors in \texttt{C13} than the
      %       others.

\section{Conclusions}
\label{sec:Conclusions}
In this paper, we have described \DeepGCC{}, a method based on deep
learning that transforms the \textit{GCC-PHAT} of two signals, emitted
from the same source and received in two microphones, into a Gaussian
function whose maximum appears at the time difference between the two
signals. We use \DeepGCC{} to estimate a smoother and more accurate
acoustic power map, as compared to that generated by the standard \SRP{}
method. We obtain the acoustic source position finding the position
where this map is maximum. In our experiments, \DeepGCC{} yields more
accurate localization than \textit{GCC-PHAT} in all cases. We also
compare \DeepGCC{} with existing deep learning methods that estimate the
source position from the microphone signals directly. Our approach is
consistently more accurate than these methods when testing and training
conditions vary significantly, which makes it better suited for deployment in real scenarios. 

As future work, our method will  be combined with sparse
denoising~\cite{velasco2016denoising} to improve localization accuracy
based on acoustic maps. We also plan to extend \DeepGCC{} to multiple
speaker localization tasks.  

      %       In what respect to future work, the generation of an acoustic map will
      %       allow us to apply denoising techniques on it, like those described in
      %       \cite{velasco2016denoising}, in order to improve the \textit{ASL}
      %       accuracy.

% \section*{Acknowledgements}

% This work has been partially supported by the Spanish Ministry of
% Economy and Competitiveness under projects HEIMDAL-UAH
% (TIN2016-75982-C2-1-R) and ARTEMISA (TIN2016-80939-R); by the University
% of Alcal\'a under project ACERCA (CCG2018/EXP-019).

      %       References should be produced using the bibtex program from suitable
      %       BiBTeX files (here: strings, refs, manuals). The IEEEbib.bst bibliography
      %       style file from IEEE produces unsorted bibliography list.
      %       -------------------------------------------------------------------------
\bibliographystyle{IEEEbib}
\bibliography{strings,paperICASSP2020-juanma,refs}

\end{document}